\documentstyle{article}

\def\1ad{\mbox{\normalsize $^1$}}
\def\2ad{\mbox{\normalsize $^2$}}
\def\3ad{\mbox{\normalsize $^3$}}
\def\4ad{\mbox{\normalsize $^4$}}
\def\5ad{\mbox{\normalsize $^5$}}
\def\6ad{\mbox{\normalsize $^6$}}
\def\7ad{\mbox{\normalsize $^7$}}
\def\8ad{\mbox{\normalsize $^8$}}
\def\makefront{\vspace*{1cm}\begin{center}
\def\newtitleline{\\ \vskip 5pt}
{\Large\bf\titleline}\\
\vskip 1truecm
{\large\bf\authors}\\
\vskip 5truemm
\addresses
\end{center}
\vskip 1truecm
{\bf Abstract:}
\abstracttext
\vskip 1truecm}
\begin {document}
\def\titleline{
Supersymmetrization of generalized 
\newtitleline
Freedman-Townsend models
}
\def\authors{
Friedemann Brandt\footnote{Supported by
the Deutsche Forschungsgemeinschaft} and Ulrich Theis
}
\def\addresses{
Institut f\"ur Theoretische Physik, Universit\"at Hannover,\\
Appelstr.\ 2, D--30167 Hannover, Germany
}
\def\abstracttext{
We review briefly generalized Freedman-Townsend models
found recently by Henneaux and Knaepen, and provide
supersymmetric versions of such models in four dimensions which
couple 2-form gauge potentials and ordinary gauge fields
in a gauge invariant and supersymmetric manner. The latter models
have the unusual feature that, in a WZ gauge,
the supersymmetry transformations do not commute with all
the gauge transformations.
}
\large
\makefront
\section{Motivation}

We report on work \cite{BT} which was motivated by
two recent developments. One of these
is the construction of new four dimensional
$N=2$ supersymmetric gauge theories \cite{N=2}
by gauging the central charge of $N=2$ vector-tensor multiplets
\cite{VT}.
The other one is a classification of possible gauge invariant
interactions of $p$-form gauge potentials \cite{HK} in
any spacetime dimension. It turned out that these two
developments overlap:
non-supersymmetric limits of the models \cite{N=2},
obtained by neglecting the fermions and freezing scalars
to constants, are contained in a class of gauge theories found in
\cite{HK}. The latter gauge theories generalize $n$-dimensional
Freedman-Townsend
models \cite{FT}: in addition to the $(n-2)$-form gauge potentials
present in Freedman-Townsend models, they contain
$p$-form gauge potentials with $p<n-2$ interacting
with each other and with the $(n-2)$-form gauge potentials.
We call such gauge theories Henneaux-Knaepen models.
Specified to four spacetime dimensions, they give interactions
between 2-form gauge potentials and ordinary gauge fields.
The $N=2$ supersymmetric models \cite{N=2} 
may thus be regarded as supersymmetric versions of special
Henneaux-Knaepen models. A natural question in this context is
whether more general Henneaux-Knaepen models can be supersymmetrized
too. Our work provides $N=1$ supersymmetric
versions of such models in four spacetime
dimensions.

\section{Interactions between $p$-form gauge potentials}

We first briefly review some of the results obtained by
Henneaux and Knaepen \cite{HK}.
Their starting point is the free action for a set of
$p$-form gauge potentials $A^a$ with various form-degrees $p_a$
in $n$ spacetime dimensions,
\begin{equation}
S^{(0)}= \frac 12\int \delta_{ab} F^a\wedge *F^b ,\quad
F^a=dA^a,\quad
A^a=\frac{1}{p_a!}\, dx^{\mu_1}\wedge\dots\wedge dx^{\mu_{p_a}}
A^a_{\mu_1\dots\mu_{p_a}}.
\label{1}\end{equation}
The free action is of course invariant under the standard
abelian gauge transformations
\begin{equation}
\delta^{(0)}_\epsilon A^a=d\epsilon^a
\label{2}\end{equation}
where $\epsilon^a$ is an arbitrary $(p_a-1)$-form.

The classification of gauge invariant interactions between the $A^a$
performed in \cite{HK} proceeds along the lines of
\cite{BH}. One seeks the possible
consistent deformations $S$ and $\delta_\epsilon$ 
of the free action and its gauge symmetries. These are
deformations
such that the deformed action is invariant under the deformed
gauge transformations,
\begin{equation}
\left.
\begin{array}{c}
S=S^{(0)}+gS^{(1)}+g^2S^{(2)}+\dots\\
\delta_\epsilon=\delta^{(0)}_\epsilon
+g\delta^{(1)}_\epsilon+g^2\delta^{(2)}_\epsilon
+\dots \end{array}\right\}\quad
 \delta_\epsilon S\stackrel{!}{=}0,
\label{3}\end{equation}
where $g$ is a continuous deformation parameter. A deformation
is called trivial if it can be removed by local field
redefinitions of the $A^a_{\mu_1\dots\mu_{p_a}}$.

To first order in $g$, (\ref{3}) imposes that $S^{(1)}$ be
$\delta^{(0)}_\epsilon$-invariant {\em on-shell}.
Hence, the general form of $S^{(1)}$ is
\begin{equation}
S^{(1)}=S^{(1)}_{\mathrm{inv}}+\hat S^{(1)},\quad
\delta^{(0)}_\epsilon S^{(1)}_{\mathrm{inv}}=0,\quad
\delta^{(0)}_\epsilon \hat S^{(1)}\approx 0
\label{4}\end{equation}
where $\approx$ denotes weak (on-shell) equality.

$S^{(1)}_{\mathrm{inv}}$ contains polynomials in the
field strengths $F^a_{\mu_1\dots\mu_{p_a+1}}$ and
derivatives thereof. Moreover it contains the possible
abelian Chern-Simons terms that can be constructed from the
$A^a$ present in the free model under study.
A remarkable result is the following.

{\bf Theorem \cite{HK}.} If
$p_a\in\{2,\dots,n-2\}\ \forall a$, then
$\hat S^{(1)}$ can be assumed
to be a linear combination of terms of the form
\begin{eqnarray}
& \int (*F^{a_1})\wedge\dots\wedge(*F^{a_r})\wedge
F^{a_{r+1}}\wedge\dots\wedge F^{a_{r+s}}\wedge A^{a_{r+s+1}}\ ,&
\label{5}\\
& r\neq 0,\quad 
\sum_{i=1}^r(n-p_{a_i}-1)+\sum_{i=r+1}^{r+s}(p_{a_i}+1)+
p_{a_{r+s+1}}=n. &
\label{6}\end{eqnarray}
In (\ref{6}), $r\neq 0$ guarantees that (\ref{5})
does not reduce to a Chern-Simons term (the latter 
is already contained in $S^{(1)}_{\mathrm{inv}}$). The second condition
in (\ref{6}) expresses the requirement that the integrand of (\ref{5})
be a volume-form. 

The above theorem does not hold in presence of 
1-form gauge potentials. Namely, then 
$\hat S^{(1)}$ can contain additional terms which cannot be
cast in the form (\ref{5}). An example is the cubic Yang-Mills
vertex $f_{abc}\int A^a\wedge A^b\wedge *F^c$
($f_{abc}=f_{[abc]}$, $p_a=1\ \forall a$)
which contains the product of two undifferentiated gauge potentials,
in contrast to (\ref{5}).

Note that (\ref{6}) imposes conditions on the
interaction terms (\ref{5}).
These conditions can be rather severe. Consider for instance
the case that all the $p$-form gauge potentials have the
same degree, i.e., $p_a=p$ for all $a$.
Then (\ref{6}) yields $(r-1)(n-2-p)+s(p+1)=(2-r)$ whose
left hand side is nonnegative for $p\leq n-2$
whereas the right hand side is negative for $r>2$.
Examining the two
remaining cases $r=1,2$, one finds that $p=n-2$, $r=2$, $s=0$ 
is the only solution with $p\not\in\{0,n\}$. 
This solution yields precisely the Freedman-Townsend vertices.
When, in addition to $(n-2)$-forms, there are also $p$-forms
with other degrees, then (\ref{6}) has further solutions
of a similar type, namely $(r,s)=(2,0)$, $p_1=n-2$, 
$p_2=p_3$ arbitrary. The corresponding vertices (\ref{5})
are given by
\begin{equation}
\int (*F^a)\wedge(*F^b)\wedge A^c,\quad
p_a=n-2,\quad p_b=p_c.
\label{8}\end{equation}
As also shown in \cite{HK}, linear combinations
$k_{abc}\int(*F^a)\wedge(*F^b)\wedge A^c$ of these
vertices can be completed to a consistent deformation of
the free theory to all orders in $g$
if the coefficients $k_{abc}$ with $p_a=p_b=p_c$
are the structure constants of an arbitrary Lie algebra $G$
(not necessarily compact)
and the other coefficients $k_{abc}$
define representation matrices of $G$.
Furthermore, the complete deformation can be elegantly constructed
in first order form by means of auxiliary 1-forms.

In four spacetime dimensions, the resulting models couple
2-form gauge potentials and ordinary gauge fields. 
Henceforth we denote the former by
$B_A$, the latter by $A^a$, and the auxiliary 1-forms
by $V^A$,
\begin{equation}
B_A=\frac 12\, dx^\mu\wedge dx^\nu B_{\mu\nu A}\ ,
\quad A^a=dx^\mu A_\mu^a\ ,\quad V^A=dx^\mu V_\mu^A\ .
\label{10}\end{equation}
In the first order formulation,
the deformed action and gauge transformations read
\begin{eqnarray}
& S=\int [-F^A(V)\wedge B_A+\frac 12\, \hat F^a\wedge *\hat F_a
          +\frac 12\, V^A\wedge *V_A] &
\label{11}\\
& \delta_\epsilon B_A=d\epsilon_A-g{f_{BA}}^C V^B\wedge\epsilon_C
         +2gT_{Ab}^a(*\hat F_a)\epsilon^b &
\label{12}\\
& \delta_\epsilon A^a= d\epsilon^a+gT_{Ab}^aV^A\epsilon^b,\quad
\delta_\epsilon V^A=0. &
\label{13}\end{eqnarray}
Here, $\epsilon_A$ and $\epsilon^a$ are arbitrary 1-forms and
0-forms respectively, and 
\begin{eqnarray}
& F^A(V)=dV^A+\frac 12\, g{f_{BC}}^A V^B\wedge V^C,\quad
\hat F^a=dA^a+gT_{Ab}^aV^A \wedge A^b &
\nonumber\\
& *\hat F_a=\frac 14 \delta_{ab} dx^\mu\wedge dx^\nu
      \varepsilon_{\mu\nu\varrho\sigma} \hat F^{\varrho\sigma b},
\quad *V_A=\frac 16\, \delta_{AB}dx^\mu\wedge dx^\nu\wedge dx^\varrho
          \varepsilon_{\mu\nu\varrho\sigma} V^{\sigma B} &
\label{14}\end{eqnarray}
where ${f_{AB}}^C$ and $T_A$ are structure constants and
representation
matrices of some Lie algebra $G$,
\begin{equation}
{f_{[AB}}^{D} {f_{C]D}}^{E} = 0,\quad
[T_A,T_B]={f_{AB}}^C T_C\ .
\label{15}\end{equation}

\section{D=4, N=1 supersymmetric Henneaux-Knaepen models}

We shall now briefly outline our method \cite{BT}
to supersymmetrize all the models defined through
(\ref{10}--\ref{15}). We first construct
a superspace version of these models, generalizing
earlier work \cite{CLL} on supersymmetric Freedman-Townsend models.
To that end we associate an appropriate
superfield with each of the forms in  (\ref{10}),
\begin{equation}
B_A\rightarrow \Psi^\alpha_A\ ,\quad
A^a\rightarrow{\cal A}^a\ ,\quad
V^A\rightarrow{\cal V}^A,
\label{16}\end{equation}
where $\Psi^\alpha_A$ is a chiral spinor-superfield as in
\cite{siegel}, while ${\cal A}^a$ and ${\cal V}^A$ are real
vector-superfields. From
${\cal A}^a$ and ${\cal V}^A$, we construct two chiral superfields,
$Y^a_\alpha$ and $W^A_\alpha$,
\begin{eqnarray} 
& Y_\alpha^a = - \frac{{\mathrm{i}}}{4} \bar D^2 ({\mathrm{e}}^{-2
{\mathrm{i}} {\cal V}} D_\alpha\, {\mathrm{e}}^{{\mathrm{i}}
{\cal V}}\!\! {\cal A})^a\ ,\quad {\cal V} = g {\cal V}^A T_A &
\label{17}\\
&  g W_\alpha^A\, T_A = - \frac{{\mathrm{i}}}{4} \bar D^2 
  ({\mathrm{e}}^{-2{\mathrm{i}} {\cal V}}\! D_\alpha\, {\mathrm{e}}^{2
{\mathrm{i}} {\cal V}})\ .&
\label{18}\end{eqnarray}
Thanks to the chirality of $\Psi^\alpha_A$, $Y_\alpha^a$ and
$W_\alpha^A$, the following Lagrangian is manifestly supersymmetric,
\begin{equation}
L= \int\! d^2 \theta\, [ W^A \Psi_A+
\delta_{ab} Y^a Y^b+d^2 \bar\theta\,\delta_{AB}{\cal V}^A{\cal V}^B]
+ \mbox{c.c.}
\label{19}\end{equation}
This Lagrangian gives indeed a supersymmetric version of
(\ref{11}). This is seen by working out
$L$ in component form and by verifying that it is
gauge invariant, up to a total derivative, 
under the following transformations
of $\Psi^\alpha_A$, ${\cal A}^a$ and ${\cal V}^A$ which
are the counterparts of the gauge transformations
(\ref{12}) and (\ref{13}):
\begin{eqnarray}
&  \delta \Psi_{\alpha A} = {\mathrm{i}} \bar D^2 ({\mathrm{e}}^{-2
{\mathrm{i}}\hat {\cal V}}\! D_\alpha\,
  {\mathrm{e}}^{{\mathrm{i}}\hat {\cal V}}\! C)_A-2{\mathrm{i}} g\,
\delta_{ab} Y^a\, T_{Ac}^{b}\, \Lambda^c,
\quad {\hat {\cal V}_A}{}^B=-g{\cal V}^C{f_{CA}}^B &
\label{20}\\
& \delta {\cal A}^a={\mathrm{i}} ({\mathrm{e}}^{{\mathrm{i}}
 {\cal V}}\! \Lambda - {\mathrm{e}}^{-{\mathrm{i}} {\cal V}}
	\bar{\Lambda})^a\ ,\quad \delta {\cal V}^A = 0. &
\label{21}\end{eqnarray}
Here $C_A$ and $\Lambda^a$ are arbitrary real vector superfields
and chiral superfields respectively
($\bar D_{\dot\alpha}\Lambda^a=0$).

Now, the gauge transformations (\ref{20}) and (\ref{21}) act as shift
symmetries on some of the component fields of the superfields $\Psi_A$
and ${\cal A}^a$.
As usual, this means that the action can actually be
written in terms of fewer fields, with a correspondingly reduced gauge
invariance and modified supersymmetry transformations. 
Such a ``WZ gauged'' version of the above models is also 
constructed in
\cite{BT}. The surviving component fields of
$\Psi_A$ are those of a real linear multiplet, i.e., 
a real scalar field $\varphi_A$, 
a real 2-form gauge potential $B_{\mu\nu A}$,
and a Weyl spinor $\chi_A$. The surviving component fields of
${\cal A}^a$ are a gauge field
$A_\mu^a$, a Weyl spinor $\lambda^a$ and a real 
auxiliary field $D^a$. The component fields of ${\cal V}^A$ are
auxiliary fields which may be eliminated using the
equations of motion. The supersymmetry
algebra closes only modulo gauge transformations. 
All these features are expected from the
experience with other supersymmetric gauge theories in the
WZ gauge, such as standard super Yang-Mills theories.
However, there is also a remarkable difference, as compared
to other supersymmetric gauge theories: in the WZ gauge
constructed in \cite{BT},
the supersymmetry transformations do not commute with all
the gauge transformations, not even on-shell!

%


\end{document}